\documentclass[acmsmall,screen,nonacm]{acmart}

\AtBeginDocument{%
  }

\setcopyright{acmlicensed}
\copyrightyear{2026}
\acmYear{2026}

\usepackage{multirow}
\usepackage{graphicx}
\usepackage{float}
\usepackage{listings}
\usepackage{listings, listings-rust}
\usepackage{makecell}

\begin{document}

\title{LLM4C2Rust: Large Language Models for Automated Memory-Safe Code Transpilation}


\author{Sarah Bedell}
\email{sarah.bedell@uga.edu}
\orcid{0009-0007-1009-9920}

\affiliation{
  \institution{Department of Computer Science \\
  University of Colorado Colorado Springs (UCCS)}
  \city{Colorado Springs}
  \state{Colorado}
  \country{USA}
}

\affiliation{
  \institution{\\ \& University of Georgia, Athens, GA}
  \city{Athens}
  \state{Georgia}
  \country{USA}
}

\author{Nazanin Siavash}
\email{nsiavash@uccs.edu}
\orcid{0009-0000-4177-0632}
\affiliation{
  \institution{Department of Computer Science \\
  University of Colorado Colorado Springs (UCCS)}
  \city{Colorado Spring}
  \state{Colorado}
  \country{USA}
}

\author{Armin Moin}
\email{amoin@uccs.edu}
\orcid{0000-0002-8484-7836}
\affiliation{
  \institution{Department of Computer Science \\
  University of Colorado Colorado Springs (UCCS)}
  \city{Colorado Spring}
  \state{Colorado}
  \country{USA}
}

\renewcommand{\shortauthors}{Bedell et al.}

\begin{abstract}
\paragraph{Abstract} Memory safety has long been a critical challenge in software engineering, particularly for legacy systems written in memory-unsafe languages, such as C and C++. Rust, one of the youngest modern programming languages, offers built-in memory-safety guarantees that make it a strong candidate for secure systems development. Consequently, transpiling C/C++ code into memory-safe Rust code has become a growing area of research. However, manual transpilation is often time-consuming and error-prone. Additionally, rule-based automated approaches are not as flexible or cost-effective as methods enabled by state-of-the-art AI models, techniques, and methods, such as those that deploy Large Language Models (LLMs), for example, Generative Pretrained Transformers (GPT). In this paper, we propose a Retrieval-Augmented Generation (RAG)–assisted framework that integrates an LLM with a Small Language Model (SLM) to perform a C/C++-to-Rust transpilation with a focus on enhancing memory safety. The framework deploys a segmentation strategy that processes C/C++ code in balanced blocks, guiding the LLM with retrieved context from Rust documentation and compiler error references. Our experiments using three OpenAI models (GPT-4o, GPT-4-Turbo, and O3-Mini) demonstrate that the RAG-enhanced pipeline generally improves both code correctness and security for C-to-Rust code transpilation. Several \textit{Coreutils} programs achieve complete elimination of Raw Pointer Dereferences (RPDs) and Unsafe Type Casts (UTCs) in the final Rust output, indicating the potential of LLM-based transpilation for advancing automated software modernization and repair, as well as memory-safe code generation.
\end{abstract}



\maketitle

\section{Introduction}\label{sec:introduction}
For decades, the C programming language has been a cornerstone of low-level software development, powering operating systems, embedded controllers, and applications where efficiency and direct hardware access are paramount. Its fine-grained control over memory and hardware registers has made it indispensable in performance-critical domains. However, this control comes at the cost of manual memory management, which frequently leads to vulnerabilities such as buffer overflows, dangling pointers, and data races. Industry studies have estimated that approximately 70\% of reported security vulnerabilities stem from these memory-safety issues \cite{Sim+2025}. The U.S. federal government has also called for a shift toward memory-safe languages to reduce such risks by design \cite{whitehouse2024ncstrategy,whitehouse2024secure}.

Rust has emerged as a modern alternative for software system programming that addresses many of these challenges by enforcing a strict \textit{ownership-and-borrowing} model at compile time, preventing entire classes of memory-safety bugs before execution. Its performance is comparable to C and C++, while its safety guarantees have led to successful adoption in high-profile projects. Despite these advantages, vast amounts of critical infrastructure remain written in C. Migrating this legacy code base to Rust in a manual manner would be both costly and time-consuming. This challenge has motivated a growing interest in automated C-to-Rust transpilation.

Existing automatic transpilation (i.e., translation across programming languages) techniques generally fall into two main categories: rule-based and modern-AI-based. The early generation of AI-enhanced systems was rule-based. While effective at maintaining functional equivalence, such methods often generate Rust code that contains unsafe blocks and retains low-level idioms from C, limiting both its maintainability and the security benefits of Rust. In contrast, state-of-the-art AI approaches, for example, based on Large Language Model (LLMs), such as Generative Pretrained Transformers (GPT), can produce more idiomatic and typically more secure Rust code, as these models learn from large corpora of human-written code. However, LLMs suffer from a lack of semantic guarantees and are prone to \textit{hallucinations}, producing code that deviates from the original intent of the program. Therefore, we explore hybrid solutions that integrate LLMs with external components to combine the fluency of generative models with the determinism of static tools.

In this study, we focus on C/C++-to-Rust transpilation to improve memory safety, leveraging Rust’s compile-time enforcements to yield inherently more secure programs. Unlike higher-level memory-safe languages, such as Python or Java, Rust also preserves near-C-level performance, making it a practical target for system-level transpilation. We aim to automate this using cutting-edge LLM-assisted methods and techniques. In particular, our approach combines an LLM (e.g., GPT-4o, GPT-4-Turbo, or o3-mini) with a Retrieval-Augmented Generation (RAG) pipeline that contextualizes the prompts sent to models with relevant code patterns and domain-specific documentation. This design aims to mitigate hallucination-related risks and enhance the idiomatic quality, correctness, and security of the generated Rust code.
Our approach also aligns with several prominent initiatives to improve software memory safety, exemplified by the DARPA TRACTOR initiative \cite{DARPA}. 

We evaluate our proposed approach using a subset of the benchmark introduced by Nitin et al. \cite{Nitin+2025}, which comprises seven diverse \textit{Coreutils} programs originally written in C along with ten existing C2Rust-translated programs. We limit the experiments to the seven Coreutils applications to make it more feasible in terms of scope. However, the dataset’s diversity in program size and functionality provides a meaningful basis for assessing transpilation quality and performance. Additionally, our exploratory interactions with the LLMs indicated a flexibility to address both C and C++ transpilation to Rust. However, we limited the scope of the experimental study in this paper to C code transpilation only.

This paper makes the following contributions:

\begin{enumerate}
    \item We propose a novel transpilation framework that integrates LLMs with a RAG-enhanced pipeline to improve the quality (including security) of C/C++-to-Rust transpilation.

    \item We assess the ability of state-of-the-art LLMs to generate memory-safe Rust code while retaining functional equivalence to the original C code.

    \item We investigate whether the common hallucination problem in LLM-based code generation can be mitigated by making the LLM more context-aware through augmenting the prompt with targeted retrieval from relevant sources.

\end{enumerate}
Specifically, we address two Research Questions (RQs):

\textbf{RQ1:} Can LLMs effectively and efficiently transpile C code to idiomatic and memory-safe code in Rust?

\textbf{RQ2:} Can a RAG-based pipeline reduce the LLM hallucinations for code generation, thus improving the correctness of the LLM-generated/transpiled Rust code?

The remainder of this paper is structured as follows. Section \ref{sec:background} provides brief background information on various topics. Section \ref{sec:related-work} reviews related work in the literature. Next, Section \ref{sec:proposed-approach} proposes our novel approach. The experimental study in Section \ref{sec:experimental-results} validates the proposed approach. Section \ref{discussion} discusses the results and answers the RQs. Furthermore, we point out a few threats to the validity of the outcomes in Section \ref{sec:threats-to-validity}. Finally, Section \ref{sec:conclusion-future-work} concludes the paper and outlines directions for future work.

\section{Background} \label{sec:background}

    \begin{figure*}[h]
        \centering
        \includegraphics[width=.8\linewidth]{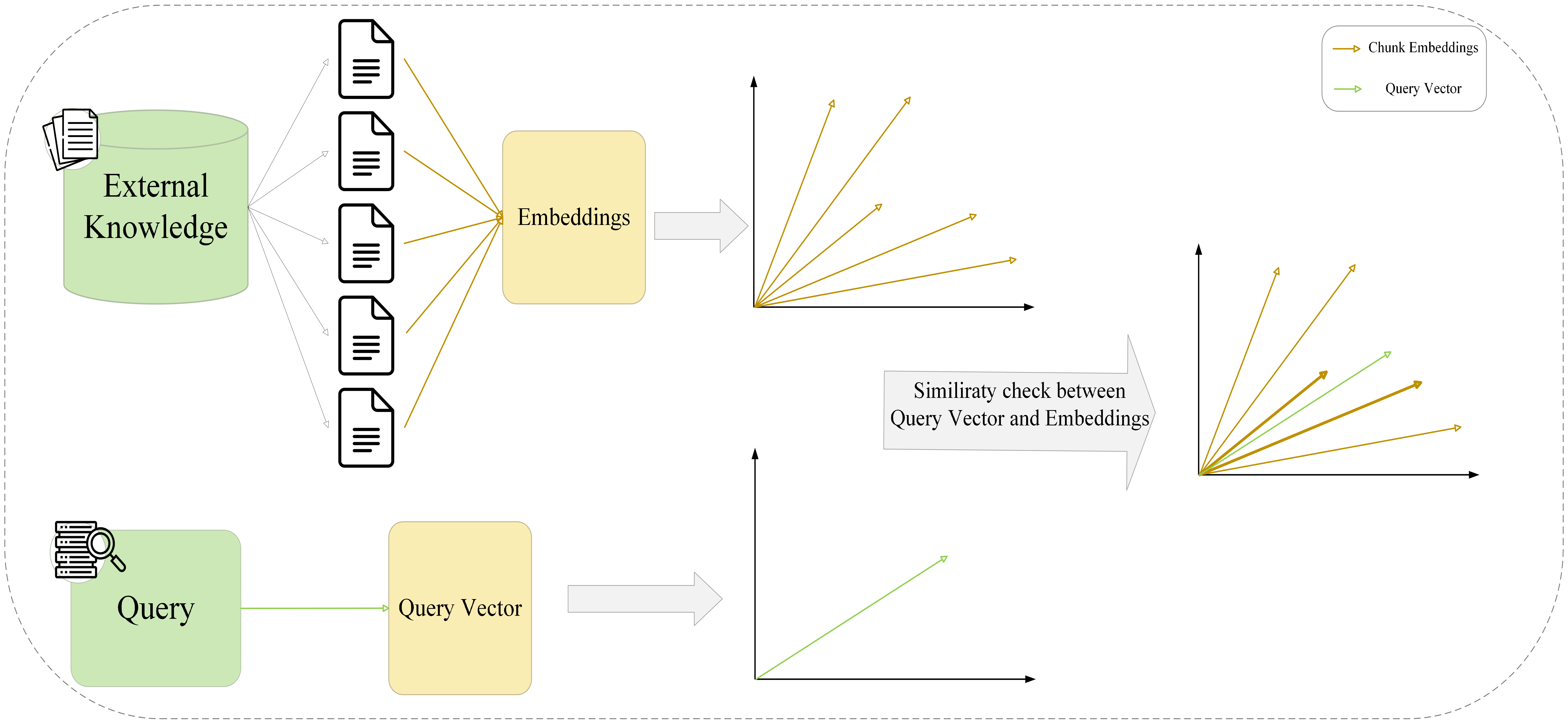}
        \caption{The RAG similarity search process. External documents are embedded into vectors, and a query is embedded into a query vector. Similarity check (e.g., Cosine Similarity) identifies the most relevant document chunks to augment the query with contextual knowledge.}
        \label{fig:ragex}
    \end{figure*}
    
    This section provides a general background of code transpilation, LLMs, RAG, memory safety, and Rust, with a central focus on how these concepts relate to the proposed approach of the paper. 

    \subsection{Code Transpilation} \label{ssec:code-transpilation}
        Code transpilation, the process of transferring source code from one programming language to another, was historically performed manually using extensive hand-crafted rules and requiring deep domain expertise \cite{Lachaux+2020}. While this approach facilitated some early efforts, it proved to be both time-consuming and resource-intensive, ultimately making it inefficient and ineffective. As a result, manual transpilation methods saw limited adoption in practice. In recent developments of software research, there has been a continuous effort to use LLMs to lower the time necessary to transpile code, yet keep the code usable, idiomatic, and efficient \cite{Roziere+2020, Bhatia+2024}.

         Code transpilation serves various purposes; however, one of the primary focuses of this paper is the improvement of safety by transpiling a program from an unsafe language, such as C or C++, to a safe language, such as Rust.

    \subsection{Large Language Models (LLMs)}   \label{ssec:llms}
        LLMs, in their recent developments, have become a large area of intrigue. While many LLMs are open source and freely available for various applications, it is often the case that user interactions contribute to ongoing training, helping improve the model’s accuracy and performance \cite{Hagos+2024}. LLMs can be applied in a wide range of tasks, including storytelling, answering questions, and generating or repairing code. However, producing usable and idiomatic code in high-level languages has historically been a significant challenge for developers working with LLMs. Recent developments have shown that many LLMs are now more proficient at generating usable code than the average developer, with many programs even offering AI assistance for code completion \cite{SVYATKOVSKIY+2019}, however there are still issues with the code being idiomatic, or legible to human developers \cite{KREBSMAZUMDAR2025}. With the increasing attention brought to LLMs, researchers now attempt to solve code transpilation through the new tool at hand \cite{Bhatia+2024}, hoping to minimize computational costs and time of translation. 
    
        However, even with the rising abilities that the LLMs have shown, they are still wildly inaccurate, in many areas, and have a tendency to \textit{hallucinate} in order to maximize efficiency. Hallucinations are shown by the LLMs conjuring up false information as a response to a prompt, creating distrust of LLMs, as well as spreading new falsities as truths \cite{li_evaluating_2025, Lei+2025}. 
    \subsection{Small Language Models (SLMs)}
Small Language Models (SLMs) have emerged as efficient alternatives to LLMs, offering several distinctive advantages. 
Due to their comparatively smaller parameter counts and training datasets, SLMs typically involve a trade-off between predictive accuracy and computational efficiency. 
However, when provided with adequate contextual information, their performance can be substantially improved. 
Recent studies have shown that SLMs combined with retrieval-augmented generation (RAG) pipelines can, in certain specialized domains, outperform general-purpose LLMs \cite{liu+2024}.

Similar to LLMs, SLMs can be either open-source or proprietary. 
Open-source models such as \textit{TinyLlama} have gained popularity owing to their accessibility and ability to be deployed locally. 
SLMs can be highly effective with added context, and have the computational efficiency from their smaller nature, making them useful for highly specified tasks \cite{lee2024}.

    \subsection{Retrieval Augmented Generation (RAG)} \label{ssec:rags}
    RAG is a technique designed to enhance the quality and relevance of responses generated by an LLM by incorporating external contextual information. In a RAG setup, the coding architect constructs a pipeline that retrieves relevant knowledge from an external source, identifies documents semantically similar to the user query, and integrates that information into the context provided to the LLM for response generation. Figure \ref{fig:ragex}  illustrates this process: the embedded query is compared with document embeddings through similarity search, and the two bolded vectors represent the most relevant document embeddings retrieved. These embeddings are then combined with the query to form an enriched context for the LLM’s response.

The external knowledge base can be tailored and segmented by the architect, allowing for task-specific and domain-focused retrieval. By providing relevant information, a RAG pipeline enables the LLM to access additional knowledge, leading to more accurate and context-aware outputs \cite{CHURCH+2024}. This focused retrieval also helps minimize “noise,” or irrelevant information, that can mislead the LLM during generation. Since such noise is a key contributor to hallucinations, there is growing interest in how effectively RAG can mitigate this issue. In other words, integrating RAG with LLMs has demonstrated measurable improvements in the quality and reliability of generated outputs \cite{OKUTAN+2024}, and it holds promise for reducing hallucination rates through context-driven refinement.



    \subsection{Memory Safety} \label{ssec:memory-safety}

       Memory safety refers to a program’s ability to prevent unauthorized or unintended access to its memory, protecting against issues such as buffer overflows, dangling pointers, heap metadata overwrites, and other forms of memory corruption \cite{Xu+2021,BergerZorn2006}. Ensuring memory safety is crucial for developers to safeguard their software systems, web applications, and distributed services. However, a large portion of legacy software is written in C or C++, languages that inherently lack built-in safety mechanisms \cite{Fiala+2023}. These languages predate many protections now common in modern languages, such as Rust, Python, and Java, such as enforced type safety, ownership models, and concurrency locks.

This gap has led to growing interest in translating C/C++ programs into safer languages like Rust. Yet, manual transpilation remains slow, error-prone, and labor-intensive. To address this challenge, recent work has explored the use of LLMs to accelerate the translation process, thereby improving both the efficiency and the security of legacy software systems.

    \subsection{Rust} \label{ssec:rust}
      Compared to older programming languages, Rust is relatively young; it was first released in 2013. Despite its modern design and extensive safety features, it remains less widespread than established high-level languages such as Python or Java. However, its adoption is rapidly increasing. Rust’s youth is, in fact, one of its advantages: it integrates contemporary programming concepts focused on memory safety, security, and reliability while maintaining computational performance comparable to C  \cite{ZHYANG+2024, Lennard+2024}. Rust can operate in both \textit{safe} and \textit{unsafe} modes \cite{Ayoun+2025}. By default, the Rust compiler (\textit{rustc}) enforces strict safety guarantees that prevent memory-related vulnerabilities. Developers must explicitly use the keyword \textit{unsafe} to bypass these protections, thereby disabling the compiler’s safety checks when low-level control is required.

\section{Related Work} \label{sec:related-work}

Research and ideas do not exist in a void, instead they are built upon one another, as ideas create pathways for new concepts to take root. Thus, this research is heavily dependent on all of the research that came before it, which was attempting to solve the same, or similar problems. There have been some programs created as a way to automate code translation, regardless of the language, such as UniTrans \cite{Yang+2024}, which increased the efficacy of the LLM attempting to aid in code transpilation.

    \subsection{Transpiling C/C++ Into Unsafe Rust}

        Several efforts have been made to automate the translation of C/C++ into unsafe Rust, motivated by the widespread use of C/C++ and the notion that unsafe Rust can later be refined into safe Rust by incorporating appropriate safety measures. Despite ongoing challenges, such as preserving the program’s semantics and resolving type mismatches between the two languages, transpiling to unsafe Rust offers a key advantage: it is significantly simpler. Rust code can be made unsafe by merely adding the unsafe keyword to functions, which is far less complex than manually enforcing all the language’s safety mechanisms and ensuring type compatibility throughout the program.
        
The first notable work in this direction was C2Rust \cite{c2rust}, which enabled automated translation from C to non-idiomatic, unsafe Rust. Building on this foundation, more recent research, such as that by Okutan \textit{et al.} \cite{OKUTAN+2024}, focused on transpiling C++ to Rust with an emphasis on syntax accuracy and execution efficiency rather than security or memory safety. This marks a significant step forward, leveraging C2Rust’s framework to improve accuracy, idiomatic expression, and performance of the generated code. However, these transpilation still fell short of providing the complete memory safety guarantees that distinguish Rust from traditional systems programming languages.

\begin{figure*}[h]
            \centering
            \includegraphics[width=1\linewidth]{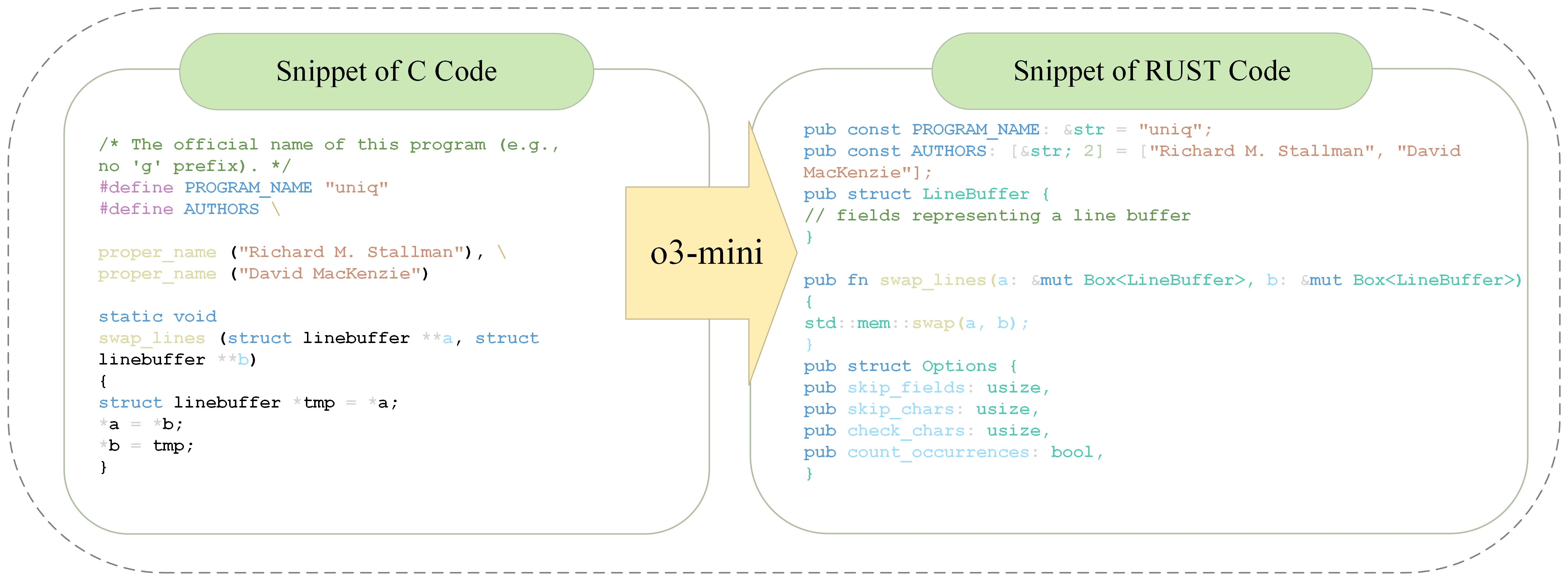}
            \caption{Transpiled C code into Rust using the o3-mini model.}
            \label{fig:tRust}
        \end{figure*}

        In Figure \ref{fig:tRust}, there's an example of C code being transpiled into Rust code. The small segment comes from the beginning of the program uniq in our dataset. This example comes from our first transpilation of the code using o3-mini. 

    \subsection{Transpiling C/C++ Into Safe Rust}
        Although further improvements are needed to enhance the idiomatic quality and semantic accuracy of the newly transpiled Rust code, the ability to generate generally compilable Rust now opens the door to advancing the memory safety and security of programs once they have been migrated from C/C++.
        This study builds upon prior research, particularly the work of \cite{Nitin+2025}, which achieved the translation of C into a safer version of Rust, improving the safety of lines of code by only 24\% and type casts by 8\%. Later studies, such as LAC2R, integrated LLMs without RAG, yielding mixed outcomes compared to C2SaferRust, especially on larger datasets \cite{Sim+2025}. The use of source-to-source transpilers has further improved program reliability by addressing API-level safety rather than focusing solely on syntax \cite{LING+2022}.

    Additionally, Hong and Ryu worked on each individual part of transpiling C into safer Rust, through locks \cite{HONGRYU2023}, union tags \cite{HONGRYU2024a}, and type casts \cite{HONGRYU2024b}. Their research provided essential background for our proposed approach, demonstrating continued progress toward fully memory-safe Rust transpilation. While our work does not address locks or union tags, it closely aligns with their investigation into type-cast safety, one of the most error-prone components of C/C++-to-Rust transpilation.

    \subsection{LLM Hallucination Mitigated by RAG}
    
Significant progress has been made in minimizing LLM hallucinations through the use of RAG pipelines. Recent studies demonstrate that integrating RAG mechanisms, rather than relying solely on traditional parametric sequence-to-sequence models, enhances accuracy in knowledge-intensive NLP tasks \cite{LEWIS+2020}. These hybrid models combine parametric and non-parametric memory to improve contextual relevance and factual grounding. Nonetheless, sequence-to-sequence architectures often remain limited in their ability to accurately model conceptual correlations, motivating recent work that applies RAG to better capture the intent of user queries or prompts \cite{Ko÷+2024}. Incorporating retrieved external knowledge helps reduce hallucinations and improve the reliability of code transpilation, as shown by \cite{OKUTAN+2024}, where RAG-enabled LLMs were employed to translate C++ into Rust. Further research explored whether generation-augmented retrieval can enhance consistency and factual accuracy even more effectively \cite{SHAO+2023}. More recent efforts examined the failure modes of both standalone LLMs and RAG-enhanced LLM pipelines, proposing the integration of knowledge graphs to address persistent issues of misinformation and contextual drift \cite{AGRAWAL+2024}.

        
        



\section{Proposed Approach} \label{sec:proposed-approach}
In this section, we first formalize the problem addressed in this work and relate it to the research questions outlined in Section \ref{sec:introduction}. We then describe the architecture of the proposed framework and explain its main components in detail.
\subsection{Problem Formalization}
Let us denote an LLM as $L$, which performs a software engineering task $M$, such as code generation, enhancement, or transpilation, in a programming language $N$. In our study, $M$ corresponds to the task of transpiling C/C++ code into Rust, i.e., $N = \{\text{C/C++}, \text{Rust}\}$.
We define $P(L, M)$ as the \textit{correctness and safety performance} of model $L$ in executing task $M$, quantified using compiler-verified metrics that capture memory-safety violations. Specifically, $P(L, M)$ depends on the number of:
\begin{itemize}
    \item Raw Pointer Dereferences (RPDs), 
    \item Unsafe Type Casts (UTCs), and 
    \item Unsafe Lines of Code (ULoCs).
\end{itemize}
A lower count of these unsafe constructs implies greater memory safety and correctness of the transpiled Rust code.

Let $I$ denote the input prompt, which represents the source C/C++ function or program to be transpiled. When this prompt is augmented with domain-specific context retrieved by a RAG pipeline, we denote the enriched input as $RAG(I)$. The corresponding safety performance of $L$ under RAG augmentation is represented as $P(L, M, RAG(I))$.
For \textbf{RQ1}, we aim to determine whether LLMs can effectively perform C/C++ $\rightarrow$ Rust transpilation such that:
\[
P(L, M, I) \text{ is maximized, i.e., unsafe constructs are minimized.}
\]
In other words, we evaluate whether $L$ can produce idiomatic and memory-safe Rust code from C/C++ inputs.

For \textbf{RQ2}, we extend the formulation to a RAG-enhanced variant of the model, denoted $L_{RAG}$, which leverages retrieval of relevant external knowledge. We hypothesize that:
\[
P(L_{RAG}, M, I) > P(L, M, I),
\]
indicating that the RAG-augmented model reduces hallucinations and improves the correctness of transpiled Rust code relative to the base LLM.

Finally, the hallucination rate $H(L, M)$, defined as the proportion of incorrect or unverifiable constructs generated by $L$, is expected to satisfy:
\[
H(L_{RAG}, M) < H(L, M).
\]
Thus, our theoretical goal is to show that retrieval-augmented prompting enhances the factual accuracy, and memory safety.
\subsection{Architecture of the Proposed Solution
}
This paper introduces a framework for LLM-assisted C/C++ to Rust transpilation with an explicit focus on memory safety rather than purely idiomatic Rust. At a high level, the approach has three main components:
\begin{itemize}
    \item a segmentation and transpilation pipeline that converts C/C++ code into (initially) unsafe Rust,
    \item a RAG pipeline that provides external context to the LLM during refinement, and
    \item a verification layer that compares LLM-reported safety improvements against compiler-verified results.
\end{itemize}

\begin{figure}[h]
        \centering
        \includegraphics[width=1\linewidth]{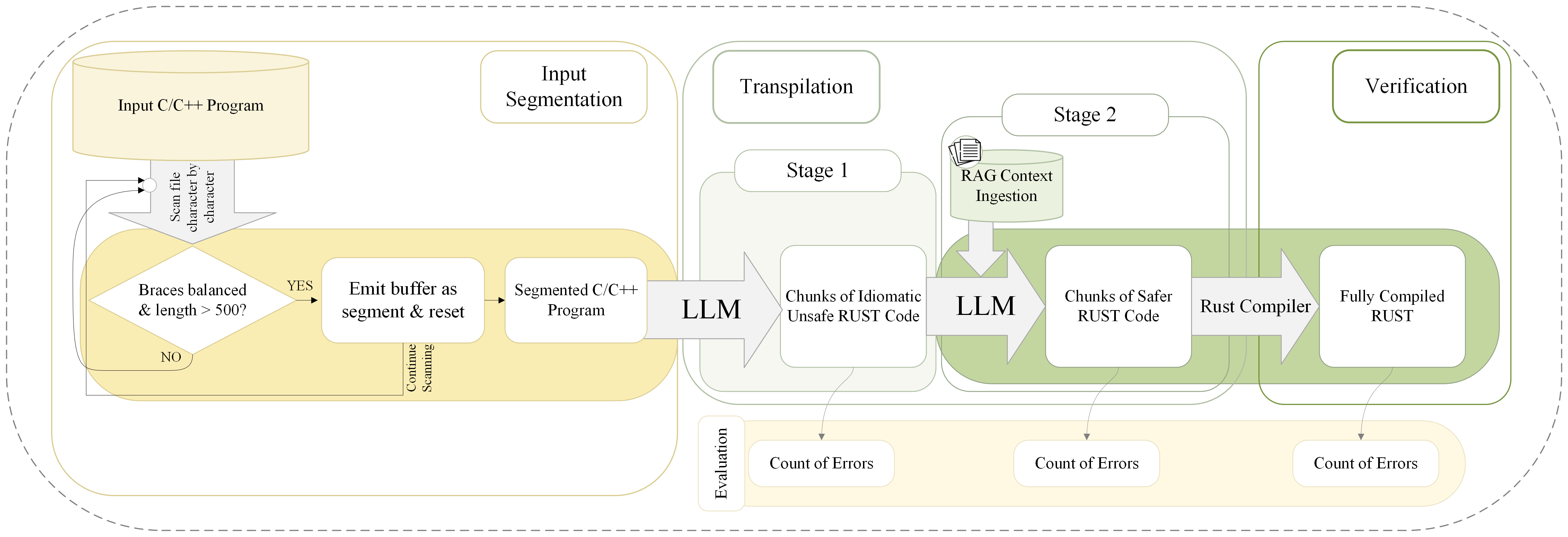}
        \caption{Comprehensive Overview of Proposed Approach}
    \label{fig:overallarchitectureofourapproach}
\end{figure}

Figure~\ref{fig:overallarchitectureofourapproach} illustrates the overall architecture of the proposed approach. We instantiate and evaluate this architecture on a subset of the dataset introduced by \cite{Nitin+2025}, consisting of seven GNU Coreutils programs written in C (\texttt{uniq}, \texttt{cat}, \texttt{pwd}, \texttt{truncate}, \texttt{head}, \texttt{split}, and \texttt{tail}). These programs span a range of code sizes and functionalities, providing a diverse set of memory-safety and control-flow patterns.

\paragraph{Input Segmentation}
Our method starts with the full C/C++ source code that we want to convert to Rust. Instead of cutting it up by individual functions, we go through the file one character at a time and keep track of every \texttt{\{} and \texttt{\}} brace. As we read, we collect the code into a temporary buffer. Whenever the braces are balanced again (meaning every \texttt{\{} has a matching \texttt{\}}) and the collected code is longer than 500 characters, we save that piece as one segment and start a new one. 

This approach ensures that the code is never split in the middle of a function or other unbalanced block. Each segment is large enough to give the LLM sufficient context (at least 500 characters) but small enough to stay within the model’s token limits. Depending on the structure of the original file, a single segment may include one large function or several smaller ones that together meet the length and balance conditions.

\begin{table}[h]
    \centering
    \caption{Amount of Segments of Code per codeutils Program}
    \begin{tabular}{|c||p{1.75cm}|p{1.5cm}|p{1.5cm}|}
        \hline
       Dataset & Program & \# of Segments & \# of LoC\\
       \hline
       \multirow{7}{*}{coreutils} 
        & uniq & 12 & 544 \\
        \cline{2-4}
         & cat & 10 & 693 \\
        \cline{2-4}
         & pwd & 12 & 333\\
        \cline{2-4}
        & truncate & 4 & 343 \\
        \cline{2-4}
        & head & 20 & 932 \\
        \cline{2-4}
        & split & 22 & 1494 \\
        \cline{2-4}
        & tail & 42 & 2205 \\
        \hline
    \end{tabular}
    \label{tab:No-of-segments}
\end{table}

Table~\ref{tab:No-of-segments} summarizes the number of segments and the original C lines of code (LoC) for each program. The segment count is driven by character length and functional structure rather than LoC alone, so we report both metrics. To further respect token limits, the system is also configured to re-segment any generated Rust file that exceeds a fixed character budget, which may slightly change the final number of segments used in later stages.

\paragraph{Two-Stage Transpilation to Safe Rust}
Before transpilation, a brief natural-language summary of each program’s overall purpose is generated and attached to every segment. This helps maintain global intent when the LLM processes each segment in isolation.

The transpilation and refinement pipeline proceeds in two main stages:
\begin{enumerate}
    \item \textbf{Initial transpilation to unsafe Rust.} Each C/C++ segment, together with the program-level summary, is sent to the LLM to produce an initial Rust version. At this stage, the model is allowed to use \texttt{unsafe} constructs, raw pointers, and casts as needed to preserve behavior.
    \item \textbf{Refinement to safer Rust.} Each initial Rust segment is then passed through a second LLM-based refinement step. In this step, the prompt explicitly instructs the model to:
    \begin{enumerate}
        \item improve general memory safety and security, and
        \item specifically reduce or eliminate raw pointer dereferences and unsafe type casts.
    \end{enumerate}
\end{enumerate}
Once all segments have been refined, they are reassembled into a single Rust file per program that preserves the functional intent of the original C/C++ program.

\paragraph{Verification}
After the initial and refined Rust versions are constructed, the framework analyzes them along two axes: (1) model-reported safety and (2) compiler-verified safety.

First, each segment is examined by the LLM to estimate the number of raw pointer dereferences (RPDs) and unsafe type casts (UTCs) in both the original and refined Rust versions. These self-reported counts provide the model’s own view of how much it believes it has improved memory safety.

Second, both versions of the compiled Rust code are passed to the Rust compiler and a custom \texttt{bash} script:
\begin{itemize}
    \item \texttt{rustc} produces error messages, from which we extract specific error codes related to RPDs and UTCs (E0133, E0392, E0793 for RPDs; E0604–E0607 for UTCs).
    \item the script scans the code for unsafe blocks (\texttt{unsafe \{...\}}) and counts unsafe lines of code (ULoCs) that may hide compilable RPDs and UTCs without triggering explicit compiler errors.
\end{itemize}
This combination allows us to (i) measure the true number of unsafe constructs remaining after refinement and (ii) compare these counts with the LLM’s self-reported values to estimate hallucination rates. Finally, we also track changes in unsafe blocks (UBs) and ULoCs to capture broader improvements in general security beyond RPDs and UTCs alone.

\paragraph{RAG Pipeline and Chunking Strategy.}
The RAG component provides the LLM with retrieval-based context, such as relevant excerpts from Rust documentation, safety best practices, and example patterns for safe pointer handling and casting. Because both the documentation and the generated programs can be large, we apply a chunking strategy driven by token and API limits.

For external documents (e.g., the Rust Programming Language book~\cite{rustprogramming}), we initially experiment with 200-character chunks and 20-character overlaps. We later increased this to 500-character chunks with 50-character overlaps to reduce the number of chunks per query, which lowered retrieval latency and token usage without degrading relevance. For large generated Rust files, we first enforce a 5000-character limit per segment; if exceeded, the file is re-chunked into 4000-character segments with a 15-character overlap. This ensures that each prompt has sufficient context while staying within OpenAI’s token constraints.

\begin{figure*}[h]
    \centering
    \includegraphics[width=1\linewidth]{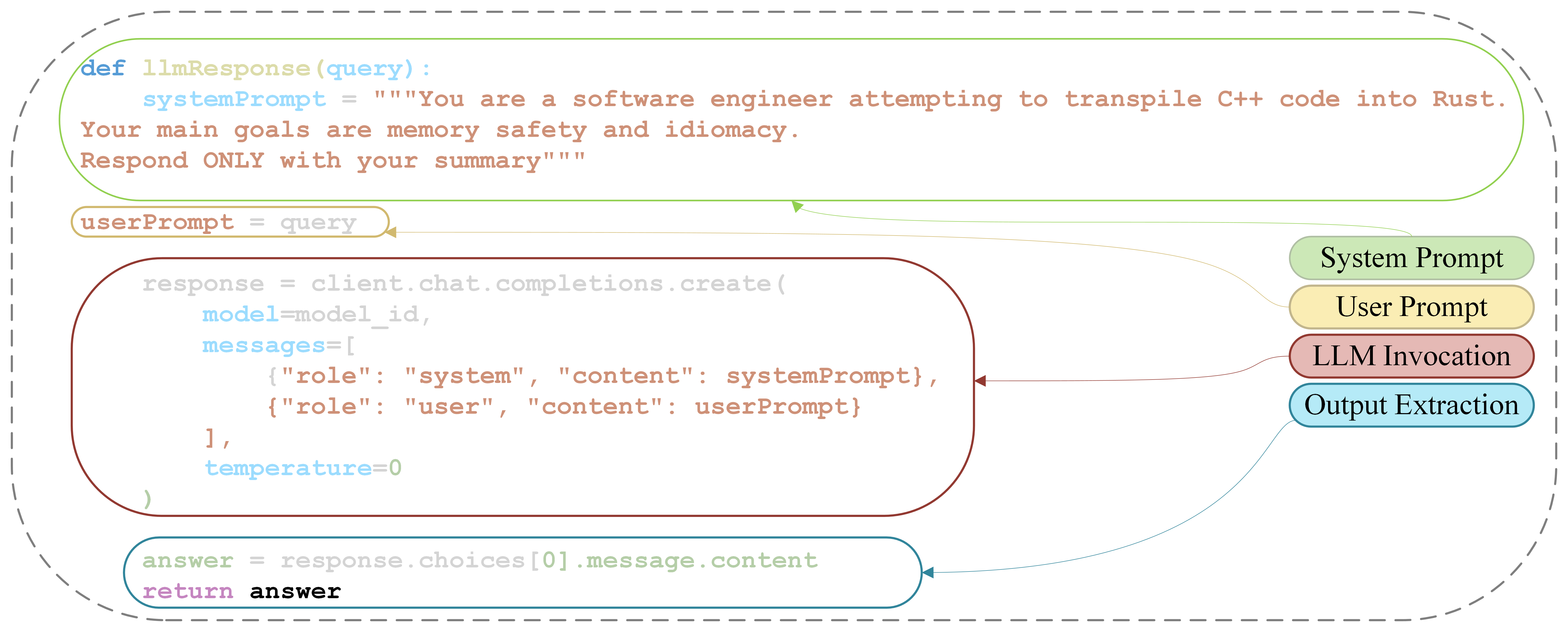}
    \caption{Annotated Python function showing the four components of the LLM-based transpilation call: (1) system prompt definition, (2) user query input (segment + summary), (3) model invocation with temperature set to zero, and (4) output extraction.}
    \label{fig:notemperature}
\end{figure*}

\paragraph{Reproducibility and Model Configuration.}
A key design goal of the framework is to maximize reproducibility of LLM outputs, despite the inherent stochasticity of generative models. To this end, all model calls are configured with \texttt{temperature = 0} and we consistently use the first returned choice (i.e., \texttt{.choices[0]}), as illustrated in Figure~\ref{fig:notemperature}. This setup reduces randomness and makes it more likely that experiments can be replicated across runs.

For this study, we employ three OpenAI models with different capacities and training focuses:
\begin{itemize}
    \item two LLMs: \texttt{GPT-4o} and \texttt{GPT-4-Turbo}, and
    \item one SLM: \texttt{o3-mini}.
\end{itemize}
Using multiple models allows us to investigate whether the observed improvements in memory safety and hallucination reduction are consistent across architectures, directly supporting RQ1 (effectiveness of LLM-based transpilation) and RQ2 (benefits of RAG-assisted code generation).

\section{Experimental Study} \label{sec:experimental-results}
This section presents the experimental study of our RAG-enhanced framework for transpiling C code into memory-safe Rust. It introduces the dataset of Coreutils programs used for testing, defines the evaluation metrics, and finally reports the experimental results comparing GPT-4o, GPT-4-Turbo, and o3-mini against compiler-verified outputs.
\subsection{Dataset}
The dataset for this study consists of seven C programs from the GNU Coreutils package \cite{coreutils}, \verb|uniq|, \verb|cat|, \verb|pwd|, \verb|truncate|, \verb|head|, \verb|split|, and \verb|tail|. These utilities were chosen because they are fundamental system-level programs that exhibit a wide range of pointer operations, type casting, and input/output handling patterns in C. Such patterns provide an effective testing ground for evaluating memory safety and type safety in cross-language transpilation tasks.

\subsection{Evaluation Metrics}
To evaluate the effectiveness of the proposed RAG-enhanced LLM transpilation framework, we use three primary quantitative metrics, RPDs, UTCs, and ULoCs, complemented by compiler-based verification and hallucination detection.

\begin{enumerate}
    \item Raw Pointer Dereferences (RPDs):  
    Count of dereference operations performed directly on raw pointers. A reduction in RPDs indicates safer memory access and improved pointer management.  

    \item Unsafe Type Casts (UTCs): 
    Instances of unsafe type conversions (e.g., \texttt{as *const \_}) that may compromise type or memory safety. The goal is to minimize or eliminate such casts in the transpiled Rust code.

    \item Unsafe Lines of Code (ULoCs):
    Total lines of code encapsulated within \texttt{unsafe \{ ... \}} blocks or functions. This provides a direct measure of how much code bypasses Rust’s compiler-enforced safety guarantees.
\end{enumerate}

Additionally, we tracked UBs and Compiler Error Codes:
\begin{itemize}
    \item RPD-related codes: \texttt{E0133}, \texttt{E0392}, \texttt{E0793}
    \item UTC-related codes: \texttt{E0604--E0607}
\end{itemize}

These were extracted using a custom \texttt{bash} script that scanned each compiled Rust program for occurrences of unsafe constructs and compiler diagnostics. This validation method ensures reliability and minimizes bias from LLM self-estimation.

To compare our approach against state-of-the-art baselines such as C2SaferRust \cite{Nitin+2025} and LAC2R \cite{Sim+2025}, we compute percentage change metrics between the Original Rust (w/o RAG) and Final Rust (w/ RAG) outputs:
\[
\text{Change (\%)} = \frac{\text{Count}_{\text{original}} - \text{Count}_{\text{final}}}{\text{Count}_{\text{original}}} \times 100
\]
A higher percentage indicates a stronger reduction in unsafe constructs and thus greater memory safety improvement.

Finally, we qualitatively assessed LLM hallucination by contrasting each model’s self-reported RPD/UTC counts with the verified compiler counts. A smaller deviation between predicted and verified counts suggests better factual grounding and reduced hallucination during code generation.

\subsection{Experimental Results}
The main goal of this experimental analysis is to assess how effectively different language models can transpile C into memory-safe Rust, and how the integration of a RAG pipeline reduces unsafe constructs and hallucinations. Each model was tested on seven Coreutils programs, and the results were measured in terms of RPDs, UTCs, and ULoCs, both as self-estimated by the models and verified through the Rust compiler
\subsubsection{GPT-4o} Table \ref{tab:rpdutc_4o} shows the changes in RPDs and UTCs for GPT-4o between the original (w/o RAG) and final Rust (w/ RAG) outputs. While GPT-4o correctly identifies overall trends in memory safety improvements, it slightly overestimates reductions in the \texttt{tail} program, where, based on the comparison between its self-reported (25) and compiler-verified (34) RPD counts, it undercounts approximately 26.5\% of the true RPDs and hallucinates a rise in UTCs. In contrast, GPT-4o performs very well on simpler programs such as \texttt{uniq}, \texttt{pwd}, and \texttt{head}, where both RPDs and UTCs are correctly reduced to zero, fully aligning with compiler verification. The model also shows accurate safety improvement trends for \texttt{cat} and \texttt{split}, where the number of unsafe constructs decreases substantially in both self-reported and verified outputs. Overall, GPT-4o captures the direction of improvement accurately and demonstrates strong qualitative reasoning even when its numeric precision varies across larger programs such as \texttt{tail}.

\begin{table*}[h]
    \centering
    \caption{Comparison of RPDs and UTCs between the initial (w/o RAG) and final (RAG-enhanced) Rust outputs generated by GPT-4o.}
    \begin{tabular}{|c||p{1.1cm}|p{1.25cm}|p{1.25cm}|p{1.25cm}|p{1.25cm}|}
        \hline
       Model & Program & \# of RPDs in Original Rust & \# of RPDs in Final Rust & \# of UTCs in Originl Rust & \# of UTCs in Final Rust \\
       \hline
       \multirow{7}{*}{GPT-4o } 
        & uniq & 0 & 0 & 3 & 0 \\
        \cline{2-6}
         & cat & 27 & 2 & 2 & 2 \\
        \cline{2-6}
         & pwd & 0 & 0 & 0 & 0 \\
        \cline{2-6}
        & truncate & 0 & 0 & 0 & 2 \\
        \cline{2-6}
        & head & 0 & 0 & 0 & 0 \\
        \cline{2-6}
        & split & 18 & 2 & 5 & 1\\
        \cline{2-6}
        & tail & 25 & 3 & 3 & 4 \\
        \hline
    \end{tabular}
    \label{tab:rpdutc_4o}
\end{table*}

\subsubsection{GPT-4-Turbo}
Table \ref{tab:rpdutc_4turbo} presents results for GPT-4-Turbo. Although GPT-4-Turbo’s self-reported counts deviate from the compiler-verified results, it achieves the most memory-safe outputs among all tested models. In particular, the model correctly identifies and eliminates nearly all RPDs and UTCs in five of the seven Coreutils programs (\texttt{uniq}, \texttt{truncate}, \texttt{head}, \texttt{split}, and \texttt{cat}), achieving full alignment with compiler verification for most of them. For \texttt{cat}, GPT-4-Turbo predicts a complete removal of RPDs (from 8 to 0) and a substantial reduction of UTCs (from 8 to 1), which closely matches the compiler’s verified counts (5 to 0 and 0 to 0, respectively). Similarly, \texttt{head} and \texttt{split} exhibit strong consistency between the model’s predicted and verified improvements, both showing near-zero unsafe constructs in the final Rust code.

The few discrepancies arise primarily in more complex programs such as \texttt{pwd} and \texttt{tail}, where GPT-4-Turbo hallucinates minor RPD increases (1~$\rightarrow$~2) or misclassifies residual unsafe casts. These deviations likely stem from the model’s limited ability to track global pointer aliasing or cross-function data flow in longer code segments. Nevertheless, the compiler results confirm that GPT-4-Turbo produces the safest overall Rust outputs, demonstrating strong structural awareness of unsafe constructs and the lowest total number of remaining RPDs and UTCs after RAG enhancement.

\begin{table*}[h]
    \centering
    \caption{Comparison of RPDs and UTCs between the initial (w/o RAG) and final (RAG-enhanced) Rust outputs generated by GPT-4-Turbo.}
    \begin{tabular}{|c||p{1.25cm}|p{1.25cm}|p{1.25cm}|p{1.25cm}|p{1.25cm}|}
        \hline
       Model & Program & \# of RPDs in Original Rust & \# of RPDs in Final Rust & \# of UTCs in Originl Rust & \# of UTCs in Final Rust \\
       \hline
       \multirow{7}{*}{GPT-4-Turbo} 
        & uniq & 0 & 0 & 0 & 0 \\
        \cline{2-6}
         & cat & 8 & 0 & 8 & 1 \\
        \cline{2-6}
         & pwd & 1 & 2 & 1 & 0 \\
        \cline{2-6}
        & truncate & 0 & 0 & 0 & 0 \\
        \cline{2-6}
        & head & 1 & 0 & 7 & 1 \\
        \cline{2-6}
        & split & 8 & 1 & 9 & 0\\
        \cline{2-6}
        & tail & 3 & 2 & 9 & 5 \\
        \hline
    \end{tabular}
    \label{tab:rpdutc_4turbo}
\end{table*}

\subsubsection{o3-mini}
Table \ref{tab:rpdutc_o3mini} displays the results for OpenAI’s o3-mini, an SLM. While o3-mini produces syntactically correct and structurally sound Rust code, it is noticeably less accurate in its estimation of memory safety metrics. The model frequently overstates the degree of improvement between its original and final outputs, showing significant discrepancies when compared to compiler-verified counts in Table~\ref{tab:rustc}.

For instance, o3-mini reports large reductions in RPDs and UTCs for \texttt{head} (16~$\rightarrow$~5 RPDs, 19~$\rightarrow$~4 UTCs) and \texttt{split} (18~$\rightarrow$~2 RPDs, 5~$\rightarrow$~1 UTC), whereas the compiler verification shows a far less pronounced improvement. In programs like \texttt{tail}, o3-mini’s hallucinations are more evident: it predicts meaningful safety gains ( 29~$\rightarrow$~14 RPDs), yet the compiler reports residual unsafe operations that remain high ( 3~$\rightarrow$~14 RPDs). This overconfidence suggests that while o3-mini can apply consistent syntactic transformations, it lacks a deep semantic understanding of memory safety rules.

One interesting observation is that o3-mini often includes additional safety directives, such as:
\begin{lstlisting}[language=Rust]
    #![forbid(unsafe_code)]
\end{lstlisting} and 
\begin{lstlisting}[language=Rust]
    #![deny(unsafe_code)]
\end{lstlisting}
These top-level compiler attributes enhance baseline safety by explicitly forbidding the compilation of unsafe constructs unless explicitly overridden. However, their presence does not necessarily imply the full removal of unsafe blocks (\texttt{unsafe \{ ... \}}) or type-cast violations. In many cases, these directives coexist with unresolved ULoCs and UBs, indicating that o3-mini attempts to enforce safety declaratively rather than eliminating unsafe constructs programmatically.

Overall, while o3-mini’s outputs demonstrate strong surface-level code quality and consistent adherence to Rust’s syntax and style conventions, the model exhibits the highest hallucination rate among the evaluated systems. Its safety improvements are often overstated, with large deviations between self-estimated and compiler-verified results. Despite these shortcomings, o3-mini’s inclusion of explicit safety directives shows a distinct behavioral difference from larger LLMs, reflecting a pattern of cautious but superficial enforcement of Rust’s memory-safety guarantees.

\begin{table*}[h]
    \centering
    \caption{Comparison of RPDs and UTCs between the initial (w/o RAG) and final (RAG-enhanced) Rust outputs generated by OpenAI's o3-mini.}
    \begin{tabular}{|c||p{1.25cm}|p{1.25cm}|p{1.25cm}|p{1.25cm}|p{1.25cm}|}
        \hline
       Model & Program & \# of RPDs in Original Rust & \# of RPDs in Final Rust & \# of UTCs in Original Rust & \# of UTCs in Final Rust \\
       \hline
       \multirow{7}{*}{o3-mini} 
        & uniq & 9 & 4 & 11 & 0 \\
        \cline{2-6}
         & cat & 7 & 9 & 9 & 9 \\
        \cline{2-6}
         & pwd & 3 & 3 & 1 & 0 \\
        \cline{2-6}
        & truncate & 3 & 1 & 9 & 0 \\
        \cline{2-6}
        & head & 16 & 5 & 19 & 4 \\
        \cline{2-6}
        & split & 18 & 2 & 5 & 1\\
        \cline{2-6}
        & tail & 29 & 14 & 36 & 15 \\
        \hline
    \end{tabular}
    \label{tab:rpdutc_o3mini}
\end{table*}

\subsubsection{Rust Compiler Verification}
To obtain ground-truth measurements of memory safety, we use \texttt{rustc}, the Rust compiler, as an external oracle. Rather than relying solely on the models' self-reported counts of RPDs and UTCs, we compile each \textit{Original Rust} (w/o RAG) and \textit{Final Rust} (w/ RAG) program and extract safety-relevant diagnostics. This allows us to quantify how many unsafe constructs remain in each version and to evaluate the extent of LLM hallucination.

We focus on two classes of compiler error codes:
\begin{itemize}
    \item \textbf{RPD-related codes:} \texttt{E0133}, \texttt{E0392}, \texttt{E0793}, which correspond to undefined behavior or violations caused by raw pointer dereferences.
    \item \textbf{UTC-related codes:} \texttt{E0604--E0607}, which capture incorrect or unsafe type conversions (e.g., invalid casts or mismatched types).
\end{itemize}

For each generated Rust program, we invoke \texttt{rustc} and run a custom \texttt{bash} script that:
\begin{enumerate}
    \item Parses the compiler output to count occurrences of the above error codes, yielding verified counts of RPDs and UTCs.
    \item Scans the source for \texttt{unsafe \{\,\ldots\,\}} blocks and unsafe functions to determine:
    \begin{itemize}
\item the number of unsafe blocks (UB), and
\item the number of unsafe lines of code (ULoCs), by counting lines syntactically enclosed within unsafe regions.
    \end{itemize}
    \item Aggregates these measures for each model (GPT-4o, GPT-4-Turbo, o3-mini), each program, and each version (Original vs. Final).
\end{enumerate}

Table~\ref{tab:rustc} summarizes the compiler-verified results. Several trends emerge:
\begin{itemize}
    \item GPT-4o achieves substantial reductions in RPDs and ULoCs for programs such as \texttt{cat}, \texttt{split}, and \texttt{tail}. For instance, \texttt{cat} goes from 29 to 0 RPDs and from 97 to 2 ULoCs, while \texttt{tail} drops from 34 to 1 RPD and from 72 to 19 ULoCs. However, small regressions remain (e.g., an additional UTC in \texttt{truncate}), illustrating that safety improvements are not uniform across all programs.
    \item GPT-4-Turbo frequently produces Final Rust code with no RPDs at all and dramatically reduced ULoCs (e.g., \texttt{split} from 114 to 1 ULoC, \texttt{cat} from 35 to 1 ULoC). Some programs (\texttt{truncate}) are fully safe in both versions, while others (\texttt{tail}) retain UTCs and ULoCs, indicating that residual unsafe regions are sometimes necessary or not fully repaired.
    \item o3-mini exhibits more mixed behavior. In some cases it reduces ULoCs (e.g., \texttt{uniq} from 120 to 10), but in others it increases RPDs, UTCs, or ULoCs (e.g., \texttt{cat} and \texttt{tail}), leading to worse safety than the Original Rust. This confirms that o3-mini is less reliable for memory-safety–oriented transpilation, even though it often inserts strict attributes like \texttt{\#![forbid(unsafe\_code)]} and \texttt{\#![deny(unsafe\_code)]}.
\end{itemize}

By contrasting Table~\ref{tab:rustc} with the self-reported counts in Tables~\ref{tab:rpdutc_4o}, \ref{tab:rpdutc_4turbo}, and \ref{tab:rpdutc_o3mini}, we can directly assess hallucination. GPT-4o and GPT-4-Turbo generally get the direction of change right (unsafe constructs decrease) but misestimate exact counts. o3-mini, by contrast, frequently reports improvements that are not reflected in the compiler diagnostics. 

\begin{table*}[h]
    \centering
    \caption{Compiler-verified counts of RPDs and UTCs in Rust code generated by different models.}
    \begin{tabular}{|p{0.4cm}||p{1.15cm}|p{1.15cm}|p{1.15cm}|p{1.15cm}|p{1.15cm}|p{1.15cm}|p{1.15cm}|p{1.15cm}|p{1.15cm}||}
    \hline
         \raisebox{-3em}{\rotatebox{90}{Model}}
 & Program & \# of RPDs in Original Rust & \# of RPDs in Final Rust & \# of UTCs in Originl Rust & \# of UTCs in Final Rust & \# of UB in Original Rust & \# of UB in Final Rust & \# of ULoC in Original Rust & \# of ULoC in Final Rust\\
       \hline
       \hline
       \multirow{7}{*}{\rotatebox{90}{GPT-4o}} 
        & uniq & 0 & 0 & 1 & 0 & 3 & 0 & 3 & 0\\
        \cline{2-10}
         & cat & 29 & 0 & 1 & 1 & 34 & 2 & 97 & 2 \\
        \cline{2-10}
         & pwd & 0 & 0 & 0 & 0 & 0 & 0 & 0 & 0 \\
        \cline{2-10}
        & truncate & 0 & 0 & 0 & 1 & 2 & 2 & 2 & 2 \\
        \cline{2-10}
        & head & 0 & 0 & 0 & 0 & 2 & 0 & 11 & 0 \\
        \cline{2-10}
        & split & 12 & 0 & 4 & 0 & 5 & 7 & 99 & 19 \\
        \cline{2-10}
        & tail & 34 & 1 & 6 & 3 & 25 & 11 & 72 & 19 \\
        \hline
        \hline
        \multirow{7}{*}{\rotatebox{90}{GPT-4-Turbo}} 
        & uniq & 0 & 0 & 0 & 0 & 1 & 0 & 3 & 0\\
        \cline{2-10}
         & cat & 5 & 0 & 0 & 0 & 10 & 1 & 35 & 1 \\
        \cline{2-10}
         & pwd & 0 & 0 & 1 & 0 & 1 & 2 & 7 & 5 \\
        \cline{2-10}
        & truncate & 0 & 0 & 0 & 0 & 0 & 0 & 0 & 0 \\
        \cline{2-10}
        & head & 0 & 0 & 1 & 0 & 11 & 1 & 19 & 1 \\
        \cline{2-10}
        & split & 0 & 0 & 3 & 0 & 15 & 1 & 114 & 1 \\
        \cline{2-10}
        & tail & 0 & 0 & 1 & 1 & 17 & 14 & 45 & 24 \\
        \hline
        \hline
        \multirow{7}{*}{\rotatebox{90}{o3-mini}} 
        & uniq & 1 & 4 & 0 & 0 & 7 & 5 & 120 & 10 \\
        \cline{2-10}
         & cat & 7 & 8 & 9 & 11 & 39 & 20 & 52 & 66\\
        \cline{2-10}
         & pwd & 0 & 3 & 0 & 0 & 0 & 4 & 0 & 12 \\
        \cline{2-10}
        & truncate & 0 & 0 & 0 & 0 & 4 & 5 & 4 & 5 \\
        \cline{2-10}
        & head & 0 & 7 & 11 & 4 & 24 & 14 & 51 & 38\\
        \cline{2-10}
        & split & 4 & 0 & 13 & 14 & 37 & 27 & 307 & 132 \\
        \cline{2-10}
        & tail & 3& 14 & 7 & 13 & 39 & 51 & 92 & 94 \\
        \hline
        
    \end{tabular}
    \label{tab:rustc}
\end{table*}

\section{Discussion} \label{discussion}

As mentioned earlier, Table~\ref{tab:rustc} presents the compiler-verified results of RPDs and UTCs detected by \texttt{rustc}, representing the ground-truth safety validation for each model’s output. This verification step ensures that the reported metrics accurately reflect the remaining unsafe constructs after compilation, independent of any self-reported LLM estimations. By cross-referencing these compiler-verified counts with the LLM-predicted values (Tables~\ref{tab:rpdutc_4o}--\ref{tab:rpdutc_o3mini}), we are able to quantitatively evaluate each model’s reliability and identify potential hallucinations in its self-assessment.

As shown in Table~\ref{tab:rustc}, the compiler results reveal that GPT-4o and GPT-4-Turbo align closely with verified safety improvements, while o3-mini exhibits larger deviations. These verified outputs serve as the foundation for calculating percentage changes between the Original Rust (non-RAG) and Final Rust (RAG-enhanced) versions, following the same methodology used in prior works such as \cite{EMRE+2021, Nitin+2025}. The earlier studies quantified safety improvement as a percentage reduction in unsafe constructs, allowing direct comparison with our RAG-enhanced approach.

By allowing the LLM to first self-evaluate its output and then verifying those results through the compiler, we are able to assess both factual accuracy and hallucination behavior. The RAG context significantly narrows the deviation between LLM-estimated and compiler-verified safety counts, demonstrating that contextual retrieval improves factual grounding. Across Tables~\ref{tab:rpdutc_4o}, \ref{tab:rpdutc_4turbo}, and \ref{tab:rpdutc_o3mini}, the differences from Table~\ref{tab:rustc} remain minimal for GPT-4o and GPT-4-Turbo, whereas o3-mini shows stronger divergence due to overgeneralization of safety patterns.

Since multiple programs achieve full memory safety, i.e., having zero RPDs and/or UTCs in both their original and final versions, we denote these cases as ``--'' in Tables~\ref{tab:comparison_rpd}, \ref{tab:comparison_utc}, and \ref{tab:comparison_ulc}. Cases marked ``!--'' represent instances where a previously safe program (\texttt{count}=0) gained new unsafe constructs in the final version, indicating a regression that cannot be expressed as a percentage, as they cannot be divided by their original count. These cases are separate from the cases with negative changes, which are only notable in the o3-mini results. Those with a negative percentage started with a positive count of the unsafe factor, which grew after the revisionary code was run. These notations preserve interpretability while highlighting meaningful transitions in program-level safety.

\begin{table*}[h]
    \centering
    \caption{Change in RPDs by Different Approaches}
    \begin{tabular}{c||c|c|c|c|c}
        Program & GPT-4o & Us GPT-4 turbo & o3-mini & C2SaferRust & LAC2R \\
        \hline
        \hline
         uniq & -- & -- & -300\% & 27\% & 60\% \\
         \hline
         cat & 100\%  & 100\% & -14\% & 24\% & 51\% \\
         \hline
         pwd & -- & -- & !-- & 24\%  & 54\% \\
         \hline
         truncate & -- & -- & -- & 19\%  & 58\% \\
         \hline
         head & -- & !-- & !-- & 21\% & 43\%
         \\
         \hline
         split & 100\% & -- & -- & 18\% & 43\% \\
         \hline
         tail & 97\% & -- & -367\% & 22\% & 27\% \\
    \end{tabular}
    \label{tab:comparison_rpd}
\end{table*}

\begin{table*}[h]
    \centering
    \caption{Change in UTCs by Different Approaches}
    \begin{tabular}{c||c|c|c|c|c}
       Program & GPT-4o& GPT-4 Turbo& o3-mini& C2SaferRust & LAC2R \\
        \hline
        \hline
         uniq & -- & -- & -- & 11\% & 48\% \\
         \hline
         cat & 0\% & -- & -22\% & 12\% & 48\% \\
         \hline
         pwd & -- & 100\% & -- & 0\% & 56\% \\
         \hline
         truncate & !-- & -- & -- & 8\% & 47\% \\
         \hline
         head & -- & 100\% & 64\% & 14\% & 48\% \\
         \hline
         split & 100\% & 100\% & -8\% & 6\% & 37\% \\
         \hline
         tail & 50\% & 0\% & -86\% & 6\% & 38\% \\
    \end{tabular}
    \label{tab:comparison_utc}
\end{table*}

\begin{table*}[h]
    \centering
    \caption{Change in ULoCs by Different Approaches}
    \begin{tabular}{c||c|c|c|c|c}
        Program & GPT-4o& GPT-4 Turbo& o3-mini& C2SaferRust & LAC2R \\
         \hline
         \hline
         uniq & 100\% & 100\% & 92\%& 28\% & 48\% \\
         \hline
         cat & 98\% & 97\% & -27\% & 26\% & 48\% \\
         \hline
         pwd & -- & 29\% & !-- & 25\% & 56\% \\
         \hline
         truncate & 0\% & -- & -25\% & 20\% & 47\% \\
         \hline
         head & 100\% & 95\% & 25\% & 25\% & 48\% \\
         \hline
         split & 81\% & 99\% & 57\% & 18\% & 37\% \\
         \hline
         tail & 74\% & 47\% & -2\% & 24\% & 38\% \\
    \end{tabular}
    \label{tab:comparison_ulc}
\end{table*}

This experimental study was designed around two main RQs as discussed in \ref{sec:introduction}: \\
\textbf{RQ1:} \textit{Can LLMs effectively and efficiently transpile C/C++ code to idiomatic and memory-safe Rust?} \\
\textbf{RQ2:} \textit{Can a RAG pipeline reduce hallucinations and improve the correctness of LLM-generated Rust code?} \\

In addressing \textbf{RQ1}, compiler-verified results (Table~\ref{tab:rustc}) and comparative performance summaries (Tables~\ref{tab:comparison_rpd}--\ref{tab:comparison_ulc}) show that GPT-4o and GPT-4-Turbo achieve significant reductions in unsafe constructs such as RPDs, UTCs, and ULoCs. Both models outperform prior approaches such as \textit{C2SaferRust} \cite{Nitin+2025} and \textit{LAC2R} \cite{Sim+2025} in achieving memory-safe and idiomatic Rust code generation, demonstrating the effectiveness of our framework in safe transpilation.

Regarding \textbf{RQ2}, the comparisons between self-reported LLM estimations (Tables~\ref{tab:rpdutc_4o}--\ref{tab:rpdutc_o3mini}) and compiler-verified results (Table~\ref{tab:rustc}) confirm that the integration of the RAG pipeline substantially reduces hallucinations and enhances factual grounding. By retrieving relevant Rust documentation and safety patterns during code generation, the RAG-enhanced LLMs exhibit improved consistency between predicted and verified safety metrics. Collectively, these findings validate that retrieval-guided LLMs not only improve memory safety but also enhance the correctness and reliability of cross-language code transpilation.

\section{Threats to Validity} \label{sec:threats-to-validity}
With the usage of LLMs, several threats to validity and reproducibility must be carefully considered. Although our framework incorporates multiple safeguards to ensure consistent and accurate results, LLMs remain stochastic systems whose outputs may vary depending on internal state, prompt phrasing, and contextual retrieval. We therefore recognize that, despite our efforts to mitigate these risks, certain limitations inherent to LLM-based approaches may still affect the validity of our findings.
\subsection{Internal Validity}
Internal validity refers to the reliability of the experimental setup and the correctness of the procedures used to measure results. 
A key threat in our study is the potential for LLM hallucination or inconsistent output generation, which can alter the observed number of RPDs, UTCs, or ULoCs. To minimize this, we adopt several safeguards:
\begin{enumerate}
    \item We employ multiple LLMs (GPT-4o, GPT-4-Turbo, and o3-mini) to cross-check outputs and ensure consistency.
    \item We set the temperature parameter to zero to reduce randomness and ensure deterministic responses across runs.
    \item We use a RAG-based similarity search to provide highly relevant context to the model, grounding its reasoning and minimizing hallucination.
    \item We conduct multi-metric validation by comparing self-reported and compiler-verified safety measures.
\end{enumerate}
Together, these steps enhance repeatability and minimize internal bias introduced by model variability.

\subsection{External Validity}
External validity concerns the generalizability of our results to other datasets, programming languages, or models. Our evaluation is conducted on seven Coreutils programs, representative but limited in scope. While these utilities cover a broad spectrum of memory and pointer operations, more complex software systems may exhibit different behaviors or safety patterns. Future replications on larger and more diverse codebases (e.g., open-source libraries or embedded systems) would be valuable to confirm the broader applicability of our approach. Moreover, results may vary with different model architectures or API versions, which evolve rapidly and may change over time.
\subsection{Construct Validity}
Construct validity refers to how well the metrics used reflect the intended concepts. Our study relies on compiler error codes (e.g., E0133, E0392, E0793 for RPDs and E0604–E0607 for UTCs) and counts of unsafe code blocks as quantitative indicators of memory safety. While these are well-established proxies, they do not capture all semantic aspects of unsafe behavior (e.g., logical data races or misuses within safe blocks). Nonetheless, these measures provide a reliable, objective approximation of safety compliance, consistent with prior work such as C2SaferRust \cite{Nitin+2025} and LAC2R \cite{Sim+2025}.
\subsection{Conclusion Validity}
Conclusion validity relates to the strength of the causal claims drawn from our results. While the RAG-enhanced framework clearly improves safety metrics and reduces hallucinations, small variations across runs or across LLMs may still occur. We mitigate this by cross-verifying compiler outputs. However, due to the probabilistic nature of LLMs, absolute determinism cannot be guaranteed, and minor deviations in future reproductions are expected.

\section{Conclusion and Future Work} \label{sec:conclusion-future-work}
Memory safety has become one of the most critical challenges in modern software engineering, with an increasing emphasis on translating unsafe C/C++ legacy code into safer alternatives such as Rust. While other memory-safe languages like Python and Java have been considered, Rust’s syntactic and structural similarity to C/C++ makes it uniquely suited for systems programming and efficient low-level performance. This work has explored whether LLMs, particularly when enhanced with RAG, can make this transpilation process more effective and less prone to hallucination.

Our approach applied a multi-stage RAG-assisted pipeline to segment C/C++ source code and iteratively prompt LLMs to: (1) transpile it into idiomatic Rust, (2) incorporate memory-safety improvements, and (3) estimate potential memory safety violations. These results have been then validated both automatically, through compiler diagnostics and bash-scripted detection of unsafe constructs, and comparatively, using state-of-the-art baselines such as C2SaferRust \cite{Nitin+2025} and LAC2R \cite{Sim+2025}.

Experimental results show that RAG-assisted LLMs generally and substantially improve the memory safety of transpiled code, outperforming prior methods in reducing RPDs and minimizing ULoCs. RAG-enhanced LLMs can not only assist but potentially surpass current automated transpilation methods in achieving memory safety without compromising idiomaticity. Furthermore, they can self-assess their safety improvements with limited hallucination, particularly when guided with targeted contextual retrieval.

For future work, we plan to extend this study to larger-scale codebases, particularly the ten LAERTES benchmark datasets, to assess scalability and generalizability. This will involve adapting our pipeline to support Rust-to-Rust reconfiguration and differential verification. Such extensions would allow us to explore whether RAG-assisted LLMs can serve as viable agents in large-scale legacy code migration and memory-safety assurance for industrial systems.

\section*{Data and Source Code Availability}
The source code is available in a GitHub repository \footnote{https://github.com/qas-lab/reu-sarah-bedell}.  

\begin{acks}
This material is based upon work supported by the U.S. National Science Foundation (NSF) under Grant No. 2349452. Any opinions, findings, conclusions, or recommendations expressed in this material are those of the authors and do not necessarily reflect the views of the NSF. In addition, in preparing this paper, we used generative AI tools and models, such as OpenAI's GPT models and Claude, to assist with the content and programming tasks.
\end{acks}

\bibliographystyle{ACM-Reference-Format}
\bibliography{refs.bib}


\end{document}